# GlueX: The Search for Gluonic Excitations at Jefferson Laboratory

Daniel S. Carman

*Ohio University, Department of Physics and Astronomy, Athens, OH 45701*

**Abstract.** One of the unanswered and most fundamental questions in physics regards the nature of the confinement mechanism of quarks and gluons in quantum chromodynamics (QCD). Exotic hybrid mesons manifest gluonic degrees of freedom and their detailed spectroscopy will provide the precision data necessary to test assumptions in lattice QCD and the specific phenomenology leading to confinement. Photoproduction is expected to be a particularly effective manner to produce exotic hybrids, however, existing data using photon beams are sparse. At Jefferson Laboratory, plans are underway by the GlueX Collaboration to use the coherent bremsstrahlung technique to produce a linearly polarized photon beam. A solenoid-based hermetic detector will be used to collect data on meson production and decays with statistics that will exceed existing photoproduction data by several orders of magnitude after the first year of running. In order to reach the ideal photon energy of 9 GeV required for these studies, the energy of the Jefferson Laboratory electron accelerator, CEBAF, will be doubled from its current maximum energy of 6 GeV to 12 GeV. The physics motivating the search and the status of the project are reviewed.

**Keywords:** Gluonic excitations, meson spectroscopy, Jefferson Laboratory

**PACS:** 12.38.Aw,12.39.Mk,14.40.-n,25.20.Lj

## MOTIVATION FOR THE STUDY OF HYBRID MESONS

The specific goal of the GlueX Collaboration at Jefferson Laboratory [1] is to better understand the detailed nature of confinement. The nature of this mechanism is one of the great mysteries of modern physics, and in order to shed light on this phenomenon, we must better understand the nature of the gluon and its role in the hadronic spectrum. Confinement within the theory of strongly interacting matter, Quantum Chromodynamics (QCD), arises from the postulate that gluons can interact among themselves and give rise to detectable signatures within the hadronic spectrum. These signatures are expected within hadrons known as hybrids, where the gluonic degree of freedom is excited. GlueX is designed to primarily focus on the study of the light quark hybrid mesons and its central purpose is to perform a detailed mapping of the spectrum of these states. It is in these systems where the gluonic degree of freedom is manifest and can provide for a more detailed understanding of the confinement mechanism in QCD.

Gluonic mesons represent a $q\bar{q}$ system in which the gluonic flux-tube contributes directly to the quantum numbers of the state. In terms of the constituent quark model, the quantum numbers of the meson are determined solely from the quark and antiquark. However QCD indicates that this simple picture is incomplete. Lattice QCD calculations predict that states with the flux-tube carrying angular momentum (called hybrids) should exist, as well as purely gluonic states (called glueballs). Modern lattice calculations for mesons show that indeed a string-like chromoelectric flux-tube forms between distant static quark charges as shown in Fig. 1a. The non-perturbative nature of the flux-tube leads to the confinement of the quarks and to the well-known linear inter-quark potential from heavy-quark confinement with $dV/dr \sim 1$ GeV/fm (see Fig. 1b). Confinement implies that an infinite energy is required to separate the quarks off to infinity. These calculations predict that the lowest lying hybrid meson states are roughly 1 GeV more massive than the conventional meson states. This provides a reference point for the mass range to which experiments must be sensitive.

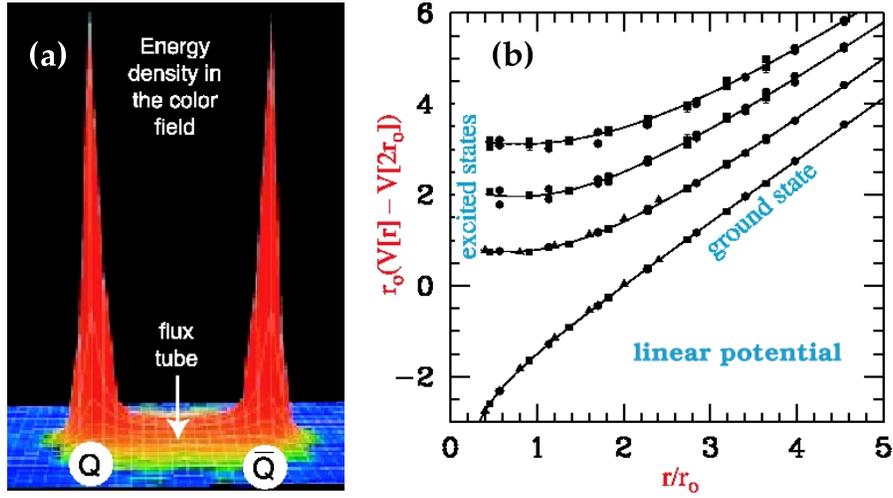

**FIGURE 1.** (a). A lattice QCD calculation of the energy density in the color field between a quark and an antiquark. This density peaks at the locations of the $q$ and $\bar{q}$ and is confined to a flux-tube stretching between the pair [2]. (b). Lattice calculation of the hybrid potential for the ground and low-lying excited states [3].

## HYBRID MESON PROPERTIES

Within the standard non-relativistic constituent quark model, conventional mesons are made up from $q\bar{q}$ pairs with the spin 1/2 quarks coupled to a total spin $S=0$ or 1. These pairs are then coupled with units of orbital angular momentum $L=0,1,2,\ldots$, and a possible radial excitation. The relevant quantum numbers to describe these states for a given principal quantum number $n$ are $J^{PC}$, where $\vec{J} = \vec{L} + \vec{S}$ represents the total angular momentum, $P = (-1)^{L+1}$ represents the intrinsic parity, and $C = (-1)^{L+S}$ represents the $C$-parity of the state.

The light-quark mesons are built up from $u$, $d$, and $s$ quarks and their antiquarks. For each value of $S$, $L$, and $n$, there are nine possible $q\bar{q}$ combinations, thus a nonet of mesons is expected for each value of $J^{PC}$. The lowest possible mass states for these conventional mesons (i.e. for $L=0$, $n=1$) are then $J^{PC} = 0^{-+}$ for $S=0$ (corresponding to the $\pi$, $\eta$, $\eta'$, and $K$ mesons) and $J^{PC} = 1^{--}$ for $S=1$ (corresponding to the $\rho$, $\omega$, $\phi$, and $K^*$ mesons). The lowest lying $S=0$ mesons have masses of about 500 MeV and the lowest lying $S=1$ mesons have masses of about 800 MeV. Nonets of higher mass mesons for a given principal quantum number are then built up by adding in units or orbital angular momentum. Conventional mesons correspond to states with the flux-tube in its ground state, and as such, the gluonic degree of freedom does not contribute.

The hybrid meson quantum numbers can be predicted within the flux-tube model [4]. In its ground state, the flux-tube carries no angular momentum. The lowest excitation is an $L=1$ rotation which contains two degenerate states (corresponding to clockwise and counterclockwise rotations). Linear combinations of these states give rise to quantum numbers of $J^{PC} = 1^{-+}$ or $1^{+-}$ for the flux-tube. Adding these quantum numbers to those for the $q\bar{q}$ pair gives the possible $J^{PC}$ for the hybrid mesons, which also are expected to have a nonet of states for each $J^{PC}$. For $S = L=0$, the possible quantum numbers are $J^{PC} = 1^{--}$ and $1^{++}$. Note that exotic hybrid mesons are *not* generated when $S=0$ as these quantum numbers are possible for conventional $q\bar{q}$ states such as $\rho$, $\omega$, and $\phi$. An important consideration in the study of these states is that non-exotic hybrids may mix with conventional $q\bar{q}$ states making clear experimental identification difficult. Establishing the hybrid nonets will depend on starting with nonets whose quantum numbers are exotic to which ordinary $q\bar{q}$ states cannot couple.

For $S=1$, $L=0$ states the possible quantum numbers are $J^{PC} = 0^{+-}$, $0^{-+}$, $1^{-+}$, $1^{+-}$, $2^{+-}$, $2^{-+}$. Of these states, those with $J^{PC} = 0^{+-}$, $1^{-+}$, and $2^{+-}$ are manifestly exotic. These unusual quantum numbers are not allowed for ordinary $q\bar{q}$ states. They provide the "smoking gun" experimental signature of gluonic excitations. Although GlueX aims to make sense of the full spectrum of light conventional mesonic states, hybrid mesons, exotic hybrid mesons, glueballs, multi-quark states, and molecular states, the exotic states will be the main initial thrust of the experimental program. The GlueX experiment has been designed to perform spectroscopy over a range of meson masses up to roughly 3 GeV.

## EVIDENCE FOR HYBRID MESONS AND PERSPECTIVES

There are two avenues to investigate in the review of current evidence for exotic hybrid mesons. On the one hand powerful indicators have been provided by current Lattice QCD (LQCD) calculations. A summary of current LQCD calculations for the lowest lying non-strange $J^{PC} = 1^{-+}$ exotic meson is provided in Table 1. These quantum numbers should correspond to the lightest nonet. These calculations are quenched or partially quenched at the current time with some employing improved lattice actions, and they indicate that the rough mass scale to search in is about 1.8 - 2.1 GeV.

On the experimental side, a number of reasonably high statistics searches for such states have been carried out with beams of hadrons and have given rise to credible evidence for the $J^{PC} = 1^{-+}$ exotics $\pi_1(1400)$, $\pi_1(1600)$, and $\pi_1(2000)$. The $\pi_1(1400)$ was first reported by the VES Collaboration at IHEP [11] and was confirmed by the E852 Collaboration at BNL in the $\pi^- p \to \eta \pi^- p$ reaction [12]. This was followed with additional confirmation by the Crystal Barrel Collaboration at CERN's LEAR facility in antiproton-neutron annihilation [13]. The strength of the exotic signal measured by Crystal Barrel was as large as that of the conventional $a_2(1230)$ meson. In the Particle Data Group (PDG) listings [14], the mass of the $\pi(1400)$ is $M=1376\pm17$ MeV and its width $\Gamma=300\pm40$ MeV with observed decays to $\pi\eta$. A second $J^{PC} = 1^{-+}$ exotic hybrid candidate, the $\pi_1(1600)$, was first observed at BNL E852 in $\pi^- p \to \rho p \to$ $\pi^+\pi^-\pi^- p$ [15]. Since that time further evidence for the $\pi_1(1600)$ has been provided through its decays into $\eta'\pi$ [16], $f_1\pi$ [17], and $b_1\pi$ [18]. Evidence in these decay modes has also been provided by VES [19]. The PDG [14] lists the mass and width of this state as $M=1596^{+25}_{-14}$ MeV and $\Gamma$ as $312^{+64}_{-24}$ MeV.

**TABLE 1.** Quenched lattice QCD calculations for the mass of the lowest-lying $J^{PC} = 1^{-+}$ non-strange exotic hybrid meson.

| Collaboration | Year | Computed Mass (GeV) | Reference |
|---|---|---|---|
| UKQCD | 1997 | 1.87±0.2 | [5] |
| MILC | 1997 | 1.97±0.09 | [6] |
| MILC | 1999 | 2.11±0.13 | [7] |
| SESAM | 1998 | 1.9±0.20 | [8] |
| Mei & Luo | 2003 | 2.013±0.026 | [9] |
| Bernard *et al.* | 2004 | 1.792±0.139 | [10] |

The interpretation of these states is presently controversial. The $\pi_1(1400)$ is significantly lighter than theoretical expectations, and its only observed decay mode into $\eta\pi$ is not expected for a hybrid. Hadronic form factors suppress those decays which couple the flux-tube orbital quantum to the relative motion of the final mesons. Thus two-body decays involving one $L=0$ and one $L=1$ meson are preferred. It has been suggested that the $\pi_1(1400)$ could represent a meson-meson molecule. Other recent work suggests that this state may actually arise from non-resonant scattering, similar to $S$-wave $\pi\pi$ scattering at low energy [20]. The same explanation can also be invoked to explain a sizeable fraction of the $\eta'\pi$ signal for the $\pi_1(1600)$ [21]. The mass of the $\pi_1(1600)$ as seen in $\rho\pi$ is still somewhat lower than theoretical expectations, but could well be a hybrid. However a recent analysis of a larger data set from BNL E852 looking for $\pi_1(1600) \to \rho\pi$ [22] using what is claimed to be an improved partial wave analysis with a larger set of basis states, claims that there is no evidence for a $J^{PC} = 1^{-+}$ meson in studies of two different $3\pi$ isospin channels. The VES results [19] provided measurements of the relative branching ratios into $b_1\pi$, $\eta'\pi$, and $\rho\pi$ as 1:1.0:1.6. These predictions are highly at odds with predictions of the flux-tube model. Thus either these three modes are not all due to a hybrid meson or there is a problem with the underlying theory.

Another candidate $J^{PC} = 1^{-+}$ meson in the literature is the $\pi_1(2000)$ which has been seen through its decays into $b_1\pi$ [23] and $f_1\pi$ [24] at BNL E852. This state currently may be the least controversial of the announced exotics in the sense that its mass is in agreement with theoretical expectations and its decays modes are fully in line with expectations for hybrid mesons. However the statistical precision of the existing data is still somewhat limited.

The current landscape makes it clear that a new dedicated spectroscopy experiment with the design and expected statistics of GlueX will be able to help greatly clarify the current ambiguous situation in the mass range below 3 GeV. This will be necessary to come to an understanding of the possible overpopulation currently reported for the $1^{-+}$ hybrid nonet, where there should be only one $\pi_1$ state.

In the experimental search for exotic hybrid mesons it is important to appreciate why exotics are hard to find. Most past studies have investigated two-body final states. It is clear now that the coupling of hybrids to low multiplicity final states is essentially non-existent. In addition, unambiguous discovery of these new states requires a detailed knowledge of the full meson spectrum and understanding of multiple decay modes. Thus a full picture of conventional mesons, hybrid mesons, exotic hybrid mesons, multiquark states, molecular states, and glueball states must be disentangled. It should also be kept in mind that many of the exotic states are expected to be relatively broad, and thus will be difficult to isolate without sophisticated analysis tools.

A key part of current experimental evidence for exotic mesons relies on sophisticated partial wave analysis (PWA) tools. This type of analysis is required to identify the $J^{PC}$ quantum numbers of the states. In simple terms, a PWA determines the production amplitudes by fitting the decay angular distributions of the final state particles. The fit includes information on the polarization of the beam and target, the spin and parity of the resonance, the spin and parity of any daughter resonances, and any relative orbital angular momentum. The analysis seeks to establish the production strengths, production mechanisms, and the relative phase motion of various production amplitudes.

While the implementation of a partial wave analysis is in principle straightforward, there are difficulties that arise due to the detector system employed as well as ambiguities within the PWA framework itself. Effects such as detector acceptance and resolution can conspire to allow strength from a dominant partial wave to appear as strength in a weaker wave. Intrinsically the PWA can be misleading if the set of basis states chosen is not complete. Both of these problems lead to a phenomenon known as "leakage" where strength that should be attributed to a specific partial wave leaks into other waves. A key part of the efforts of the GlueX Collaboration is to improve the analysis tools to reduce these potential problems. Of course designing the detector system to provide full acceptance for both charged and neutral particle final states is a requisite part of the success of the analysis, as is the ability to acquire large statistics data sets. In the next generation experiment, it is apparent that sophisticated and modern analysis tools leading to detailed spectroscopy will be the answer toward making progress in understanding the nature of confinement.

## THE NEXT GENERATION EXPERIMENT

The specifications for the next generation experiment in meson spectroscopy are already quite clear. The essential ingredients are that the detector should be hermetic for both charged and neutral particle final states, with excellent resolution and particle identification capability. This is essential for successful and unambiguous partial wave analysis. The beam energy must be high enough to allow for sufficient phase space for the production of the exotic states, however it should not be too high, otherwise the cross section for background processes will increase and overwhelm the signals that are being sought. The partial wave analysis requires high statistics experiments with sensitivity to production cross sections at the sub-nanobarn level.

In terms of beam properties, it is clear that the next generation experiment should be carried out with beams of linearly polarized photons. Photon beams are expected to be particularly favorable for the production of exotic hybrids, as the photon sometimes behaves as a virtual vector meson. When the flux tube in this $S=1$ system is excited, both ordinary and exotic $J^{PC}$ are possible. In contrast, for an $S=0$ probe (e.g. pions or kaons), the exotic combinations are not generated. To date, almost all meson spectroscopy experiments in the light quark sector have been done with incident pion, kaon, or proton probes. High flux photon beams of sufficient quality and energy have not been available, so there are virtually no data on the photoproduction of mesons with masses below 3 GeV. Thus up to now, experimenters have not been able to search for exotic hybrid mesons precisely where they are expected. Theoretical calculations indicate that the photoproduction cross sections of light quark exotic mesons should be comparable to those for conventional meson states in this energy range [25].

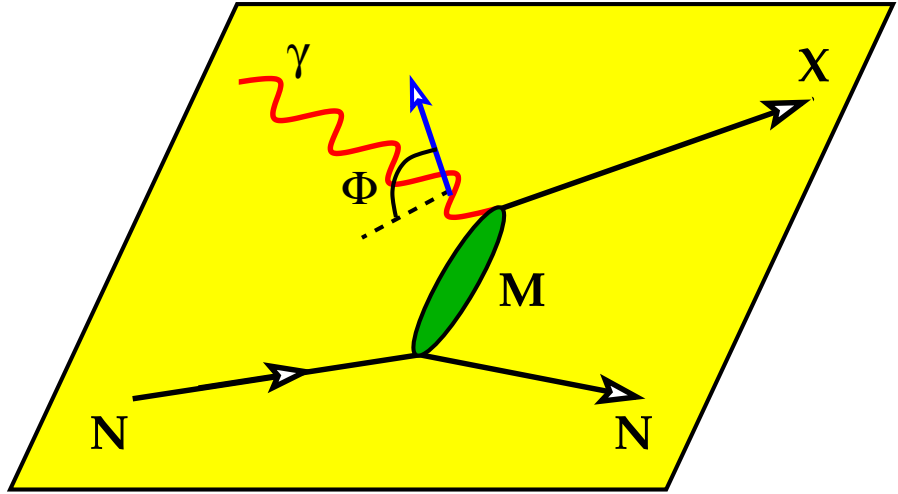

**FIGURE 2.** Generic meson exchange diagram using a photoproduction reaction with a linearly polarized photon beam showing the angle Φ used to determine the naturality of the meson *M* in the intermediate state.

As mentioned above, the next generation experiment should employ a linearly polarized photon beam. The diffractive production of a meson *X* (see Fig. 2) takes place via natural parity exchange ($J^P = 0^+, 1^-, 2^+, ...$) in the intermediate state, whereas exotic meson production takes place via unnatural parity exchange ($J^P = 0^-, 1^+, 2^-, ...$). Experiments that take place with an unpolarized or a circularly polarized photon beam cannot distinguish between the naturality of the exchanged meson *M*. However with a longitudinally polarized beam, one can distinguish the naturality of the intermediate state by selection based on the angle Φ the polarization vector makes with the hadronic production plane, where the angle is related to the naturality of the exchanged meson. This capability will be essential in helping to isolate the exotic waves in the data analysis.

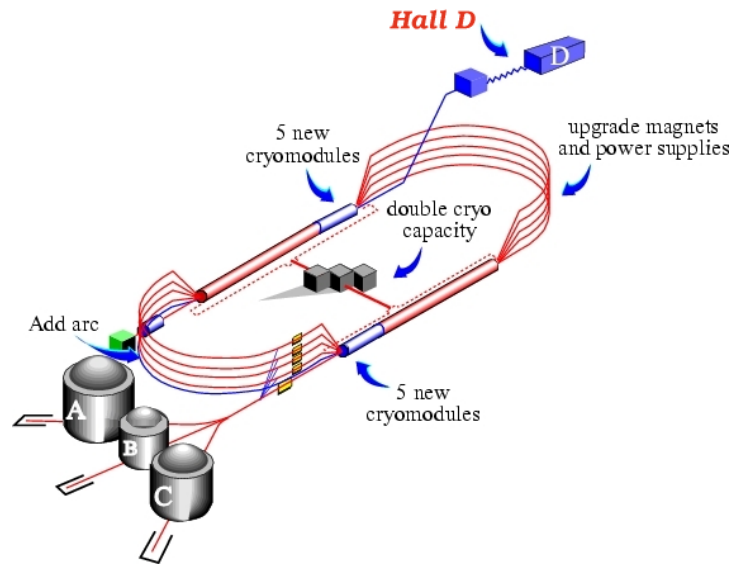

**FIGURE 3.** The configuration for the proposed 12 GeV energy upgrade of the CEBAF accelerator at Jefferson Laboratory. Also shown is the location of the new experimental Hall D for the GlueX experiment.

## THE JLAB UPGRADE & THE DESIGN OF THE GLUEX EXPERIMENT

Jefferson Laboratory represents the world's most powerful electron microscope. At its current operating point, the accelerator can deliver electron beams of up to nearly 6 GeV at currents of $\sim$200 $\mu$A and linear polarization up to $\sim$80%. The planned upgrade of this facility is designed to increase the maximum electron beam energy to 12 GeV, to construct a new experimental Hall, called Hall D, for the GlueX experiment, and to lead to equipment enhancements in the three existing Halls A, B, and C. These details are shown in Fig. 3. In April of 2004 the laboratory was awarded CD-0 approval from the U.S. Department of Energy in a first step to approve the upgrade plans. In July of 2005 the laboratory completed the next level of review that will lead to the awarding of CD-1 before the end of 2005. The current schedule has the GlueX experiment on track to begin its physics program in 2011-2012.

Fig. 4 shows a schematic layout of the planned hermetic detector to be installed in Hall D to study the photoproduction of mesons [26]. The photon beam is incident on a liquid-hydrogen target that is located within a large 2 T superconducting solenoidal magnet. The target is surrounded by drift chambers for charged particle tracking. The tracking system includes a straw tube chamber for tracking at central and backward angles and a series of cathode strip chambers for tracking at forward angles. These systems are designed to reconstruct 3–D track points with a resolution of better than 200 $\mu$m to enable accurate linking of the track hits along the helical paths followed in the magnetic field. Both of these packages will also provide for particle identification via $dE/dX$ measurements. Surrounding the tracking detectors is a barrel calorimeter for particle identi-fication and timing. Filling the upstream end of the solenoid will be another calorimeter. Downstream of the magnet is a large time-of-flight array, a Cherenkov detector for particle identification (presently planned to be a high pressure gas detector or a DIRC design), and a large lead-glass array that forms an electromagnetic calorimeter. The solenoidal geometry is ideally suited for a high-flux photon beam. The electromagnetic charged particle background (electron-positron pairs) from interactions in the target is contained along the beam line by the axial field of the magnet. The GlueX spectrometer has been designed for large and uniform acceptance for both charged and neutral particle final states, while providing the requisite resolution and particle identification capabilities for spectroscopy. The momentum resolution will be roughly $\Delta p/p \sim 1\%$.

The GlueX photon beam will be produced using the coherent bremsstrahlung technique. Here the 12 GeV electron beam will impinge on a thin ($\sim 20$ $\mu$m) diamond crystal and will produce a linearly polarized beam of 9 GeV photons after collimation. In this facility at special settings for the orientation of the crystal, the atoms of the crystal can be made to recoil together from the radiating electron leading to an enhanced emission at particular photon energies, which gives rise to linearly polarized photons. At the target the photon will achieve an average degree of linear polarization in the coherent peak of roughly 40%. In addition the photons emitted in a 0.5 GeV-wide window will be tagged using a focal plane spectrometer. The energy resolution of the photon tagger is 0.1%. This detector system, along with the high duty factor of the electron beam from the accelerator make the search for hybrid mesons quite feasible. Current plans call for the initial operation of GlueX with photon fluxes of $10^7$ s$^{-1}$, increasing to $10^8$ s$^{-1}$ when the experiment is in its nominal data acquisition mode. GlueX will collect enough data in its first year of operation to exceed existing photoproduction data by several orders of magnitude.

Hand in hand with the design of the experimental equipment, the GlueX Collaboration is performing detailed Monte Carlo studies of the physics, not only to test and improve the design of the spectrometer, but also to develop the software tools essential for the success of the data analysis. The performance of the detector, the beam flux, and the linear polarization of the photon beam determine the level of sensitivity for mapping the hybrid spectrum. The detector acceptance has been designed to be high and uniform in the relevant meson decay angles in the Gottfried-Jackson frame for *both* charged and neutral particle final states. A program of double-blind Monte Carlo studies have been carried out using the GlueX partial wave analysis software. These studies have shown that low-level exotic signals (at the few percent level) can be successfully pulled out from our data, even with relatively low statistics by GlueX standards (i.e. with only several days of running under nominal conditions). Even with extreme distortions in these simulations, leakage effects due to the spectrometer are no more than 1%. To further certify our PWA codes, consistency checks with Monte Carlo and real data will be made among different final states for the same decay mode.

At the current time the GlueX Collaboration, which has been in existence for more than 7 years, consists of more than 100 physicists from more than 25 institutions. The collaboration is growing and is very driven in its quest to construct the detector on the appropriate time line. Presently our groups are very active in refurbishing existing equipment, as well as with prototyping and designing the systems that will ultimately ensure the long term success of the experiment. The superconducting solenoid to be used as the basis for the GlueX experiment was originally part of the LASS experiment at SLAC and the MEGA experiment at Los Alamos. Presently the detector is being refurbished at Indiana University. The downstream lead-glass array and time-of-flight wall (which were originally part the BNL E852 experiment) are now being refurbished and reconfigured for GlueX. Research and development programs being lead by collaboration members at Carnegie Mellon University and Ohio University for the straw tube chambers and cathode strip chambers, respectively, are moving toward making decisions for the final tracking system design. The central calorimeter and the upstream calorimeter are nearing the end of their respective design phases at the University of Regina and Florida State University, respectively. Work is ongoing by collaborators at Indiana University and Oak Ridge National Laboratory to perform simulations of various Cherenkov options to allow the collaboration to finalize this technology choice. The collaboration also includes a number of electronics engineers and designers working at Jefferson Laboratory, Indiana University, and the University of Alberta to complete the experiment-specific electronics components (amplifiers, ADCs, TDCs, etc.) that will be necessary to meet to requirements of the GlueX experiment. The design of the tagging system by collaborators from the University of Glasgow, Catholic University, and the University of Connecticut is nearing completion and much research and development of the thin diamond radiators has taken place. A number of subsystem prototypes are presently undergoing beam tests at facilities around the world. All in all there is a lot of ongoing activity within the collaboration.

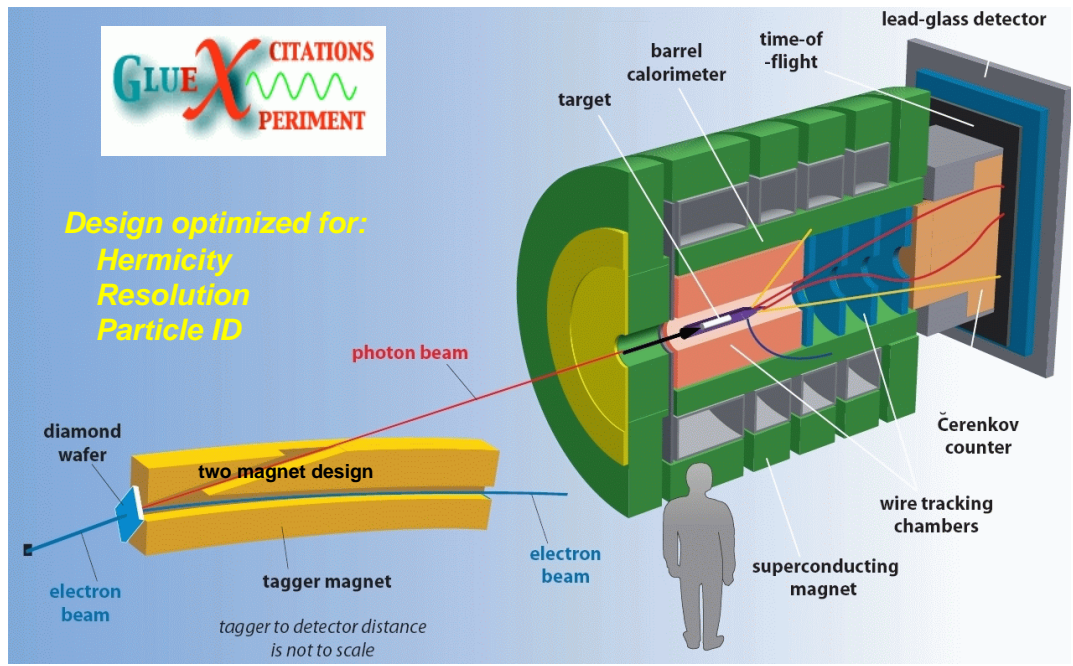

**FIGURE 4.** Schematic picture of the proposed GlueX detector and coherent bremsstrahlung photon tagging facility. The subsystems within the solenoid are indicated.

# SUMMARY AND CONCLUSIONS

Understanding confinement requires an understanding of the glue that binds quarks into hadrons. Hybrid mesons are perhaps the most promising laboratory for studying the nature of the glue. Here photoproduction promises to be rich in hybrids, starting with those having exotic quantum numbers where little or no data exist. The planned GlueX experiment that will take place at the energy-upgraded Jefferson Laboratory, will employ photon beams of the necessary flux, duty factor, and polarization, along with an optimized state-of-the-art detector. This experiment will provide for the detailed spectroscopy necessary to map out the hybrid meson spectrum, which is essential for an understanding of the confinement mechanism and the nature of the gluon in QCD.

# ACKNOWLEDGMENTS

The author thanks the conference organizers for the opportunity to discuss the GlueX experiment at HADRON05 on behalf of the GlueX Collaboration. He also gratefully acknowledges his fellow GlueX collaborators for their support and many efforts on the experiment to this point in time. This research is supported by the National Science Foundation and the U.S. Department of Energy.